\documentclass[letterpaper, 10pt, conference]{ieeeconf} 
\IEEEoverridecommandlockouts 
\usepackage{setspace, epsfig, psfrag, amssymb, amsmath, cite, color}
\usepackage{epstopdf,algorithm, algcompatible, lipsum, graphicx}

\algnewcommand\INPUT{\item[\textbf{Initial Execution:}]}
\algnewcommand\OUTPUT{\item[\textbf{Repetitive Execution:}]}
\definecolor{m1}{cmyk}{0, 0.99, 0.4429, 0.3412} 
\definecolor{m2}{cmyk}{0, 0.61, 0.4429, 0.2412} 
\definecolor{m3}{cmyk}{.83, 0, .90, .35} 
\definecolor{m4}{cmyk}{.83, 0, .40, .11} 
\definecolor{m4}{cmyk}{.83, 0, .40, .11} 
\definecolor{m5}{cmyk}{0,0.5,1,0} 
\def\bf{\normalfont\bfseries}
\def\bf{\normalfont\bfseries} 

\newfont{\smalll}{cmr8}
\def\IR{\mathbb{R}}

\def\span{\mathrm{span}}

\def\IC{\hbox{C\hskip-
.5em\raise.5ex\hbox{$\scriptscriptstyle\mid$}}\ }
\def\Ic{\hbox{\smalll C\hskip-
.5em\raise.3ex\hbox{$\scriptscriptstyle\mid$}}\ }

\def\T={\buildrel {\scriptscriptstyle\triangle} \over =}

\def\sqr#1#2{{\vcenter{\vbox{\hrule height.#2pt\hbox{\vrule
width.#2pt height#1pt \kern#1pt\vrule width.#2pt}\hrule
height.#2pt}}}}

\def\block-diag{\mathop{\rm block{\scriptstyle -}diag}}

\def\pmbb#1{\setbox0=\hbox{#1}\raise 0.5ex\box0}

\newcommand{\bequ}{\begin{eqnarray}}
\newcommand{\eequ}{\end{eqnarray}}
\newcommand{\mT}{^\mathrm{T}}

\newcommand{\rom}{\mathrm}

\newcommand {\beq}      {\begin{equation}}
\newcommand {\eeq}      {\end{equation}}

\def\IR{{\mathbb R}}
\def\IC{{\mathbb C}}

\renewcommand\theenumi{\roman{enumi}}
\renewcommand\theenumii{\alph{enumii}}
\renewcommand\theenumiii{\arabic{enumiii}}
\renewcommand\theenumiv{\Alph{enumiv}}

\def\sc{s_{\rm c}}

\def\sc{s_{\rm c}}

\definecolor{tBlue}{RGB}{25,100,250}

\definecolor{tRed}{RGB}{250,5,40}

\begin{document}
{
\title{{
{\bf Symbiotic\hspace{0.02cm} Control\hspace{0.02cm} of\hspace{0.02cm} Uncertain\hspace{0.02cm} Dynamical\hspace{0.02cm} Systems:\\Harnessing\hspace{-0.02cm} Synergy\hspace{-0.02cm} Between\hspace{-0.02cm} Fixed-Gain\hspace{-0.02cm} Control\\and\hspace{0.02cm} Adaptive\hspace{0.02cm} Learning\hspace{0.02cm} Architectures$^\star$}}}
} 

\author{Tansel Yucelen$^\dagger$, Selahattin Burak Sarsilmaz$^\ddagger$, and Emre Yildirim$^\dagger$
\thanks{$^\star$This research was supported by the United States Army Research Laboratory under the Grant {\tt{W911NF-23-S-0001.}}}
\thanks{$^\dagger$Tansel Yucelen and Emre Yildirim are with the Department of Mechanical Engineering and the Laboratory for Autonomy, Control, Information, and Systems (LACIS, {\tt\footnotesize http://lacis.eng.usf.edu/}), University of South Florida, Tampa, Florida 33620, United States of America (emails: {\tt\footnotesize yucelen@usf.edu, \tt\footnotesize emreyildirim@usf.edu}).}
\thanks{$^\ddagger$Selahattin Burak Sarsilmaz is with the Department of Electrical and Computer Engineering, Utah State University, Logan, Utah 84322, United States of America (email: {\tt\footnotesize burak.sarsilmaz@usu.edu}).}
}

\markboth{} {Shell \MakeLowercase{\textit{et al.}}: Bare Demo of IEEEtran.cls for Journals} \newcommand{\eqnref}[1]{(\ref{#1})}
\newcommand{\class}[1]{\texttt{#1}} 
\newcommand{\package}[1]{\texttt{#1}} 
\newcommand{\file}[1]{\texttt{#1}} 
\newcommand{\BibTeX}{\textsc{Bib}\TeX}
\maketitle



\begin{abstract}

Both fixed-gain control and adaptive learning architectures aim to mitigate the effects of uncertainties. 
In particular, fixed-gain control offers more predictable closed-loop system behavior but requires the knowledge of uncertainty bounds. 
In contrast, while adaptive learning does not necessarily require such knowledge, it often results in less predictable closed-loop system behavior compared to fixed-gain control. 
To this end, this paper presents a novel symbiotic control framework that offers the strengths of fixed-gain control and adaptive learning architectures. 
Specifically, this framework synergistically integrates these architectures to mitigate the effects of uncertainties in a more predictable manner as compared to adaptive learning alone and it does not require any knowledge on such uncertainties. 
Both parametric and nonparametric uncertainties are considered, where we utilize neural networks to approximate the unknown uncertainty basis for the latter case. 
Counterintuitively, the proposed framework has the ability to achieve a desired level of closed-loop system behavior even with an insufficient number of neurons (e.g., when the neural network approximation error is large) or in the face of injudiciously selected adaptive learning parameters (e.g., high leakage term parameters).

\end{abstract}  

\section{Introduction}\label{introduction}

As dynamical systems evolve in complexity, the discrepancies between mathematical models and actual physical systems continue to increase due to idealized assumptions, simplifications, linearization, degraded modes of operation, and changes in equations of motion. 
In turn, these discrepancies called uncertainties can lead to poor closed-loop system behavior and even instability. 
To mitigate the effects of uncertainties, there are two fundamental approaches in the literature; namely, fixed-gain control and adaptive learning architectures. 
For example, robust control [\citen{zhou1998essentials},\citen{yedavalli2014robust}] and sliding mode control [\citen{perruquetti2002sliding},\citen{azar2015advances}] are well-known fixed-gain control approaches, whereas adaptive control [\citen{Lavretsky},\citen{yucelenMRAC},\citen{annaswamy2021historical}] and reinforcement learning [\citen{lewis2009reinforcement},\citen{lewis2013reinforcement}] are well-known adaptive learning approaches. 

As compared to adaptive learning, fixed-gain control offers more predictable closed-loop system behavior. 
Because, the gains of the resulting control algorithm do not vary as a function of time or state. 
Yet, it requires the knowledge of uncertainty bounds for guaranteeing closed-loop system stability (see [\citen{yedavalli2014robust}, Chapter 2] and [\citen{de2016new}, Assumption 3.1] for examples). 
For dynamical systems of complex nature or when these systems experience changes in dynamics, obtaining such bounds can be challenging. 
In contrast, adaptive learning does not necessarily require such knowledge. 
However, due to their nonlinear parameter adjustment mechanism and the need for neural networks for approximating uncertainties of nonparametric nature, it often results in less predictable closed-loop system behavior compared to fixed-gain control especially during their transient period [\citen{yucelen2012low},\citen{stepanyan2012adaptive},\citen{gibson2013adaptive},\citen{yucelen2013new},\citen{gruenwald2015transient},\citen{yang2015transient},\citen{gruenwald2017direct},\citen{yang2020robust}]. 

\vspace{-0.20cm}


\subsection{Contribution and Relevant Literature}\label{contribution}

\vspace{-0.15cm}

The contribution of this paper is a novel control framework that offers the strengths of fixed-gain control and adaptive learning architectures. 
Inspired by biology, we term this framework symbiotic control since symbiosis refers to the relationship or interaction between two dissimilar organisms\footnote{Within the context of this paper, two dissimilar organisms refer to the fixed-gain control and adaptive learning architectures.} \citen{Symbiosis}. 
Specifically, symbiotic control synergistically integrates these architectures to mitigate the effects of uncertainties in a more predictable manner as compared to adaptive learning alone and it does not require any knowledge on such uncertainties. 
Both parametric and nonparametric uncertainties are considered, where we utilize neural networks to approximate the unknown uncertainty basis for the latter case. 

In the adaptive learning literature, it is well-known by the Weierstrass approximation theorem \citen{hornik1990universal} that an insufficient number of neurons can result in a large neural network approximation error over a compact region. 
Since this often leads to poor closed-loop system behavior, the authors of [\citen{joshi2019deep},\citen{le2021real},\citen{patil2021lyapunov}] have recently explored deep neural network methods to minimize this approximation error. 
In addition, it is also well-known that high leakage term parameters used in the parameter adjustment mechanisms can result in poor closed-loop system behavior since they slow down the learning process. 
Counterintuitively, the proposed symbiotic control framework has the ability to achieve a desired-level of closed-loop system behavior even with an insufficient number of neurons and without a deep neural network method, or in the face of high leakage term parameters. 

Finally, the authors of [\citen{yucelen2013new},\citen{gruenwald2015transient},\citen{gruenwald2017direct}] propose symbiotic control frameworks that are related to the results presented in this paper. 
To this end, the following two points are needed to be mentioned. 
First, none of these results focus on nonparametric uncertainties as opposed to the results documented in this paper. 
Second, although nonparametric uncertainties are not considered, the findings in \citen{gruenwald2017direct} align more closely with the results of this paper. 
However, the authors of \citen{gruenwald2017direct} make an assumption that requires some knowledge of uncertainty bounds for guaranteeing closed-loop system stability (i.e., [\citen{gruenwald2017direct}, (34)], where we here remove this assumption for both parametric and nonparametric uncertainty cases. 


\subsection{Content and Notation}\label{notation}

In the remainder of this paper, we state the problem formulation and the relevant preliminaries on fixed-gain control and adaptive learning architectures in Section \ref{problem_formulation}. 
We next present the symbiotic control framework in Section \ref{parametric_uncertainty} for dynamical systems with parametric uncertainty and in Section \ref{nonparametric_uncertainty} for dynamical systems with nonparametric uncertainty. 
Illustrative numerical examples are then given in Section \ref{numerical_examples} to show the efficacy of the contributions of this paper. 
Finally, conclusions are summarized in Section \ref{conclusion}. 

A fairly standard notation is used in this paper. 
In particular, 
$\IR$, $\IR^n$, and $\IR^{n \times m}$ respectively denote the sets of real numbers, real vectors, and real matrices; 
$\IR_{+}$, $\overline{\IR}_{+}$, $\IR^{n \times n}_{+}$, and $\overline{\IR}^{n \times n}_{+}$ respectively denote the sets of positive real numbers, nonnegative real numbers, positive-definite real matrices, and nonnegative-definite real matrices; 
and ``$\triangleq$'' denotes the equality by definition. 
Furthermore, we use 
$(\cdot)^{-1}$ for the inverse, 
$(\cdot)\mT$ for the transpose, 
$\rom{tr}(\cdot)$ for the trace, 
$\left\|\cdot\right\|_{2}$ for the vector Euclidean norm or the matrix induced 2-norm, 
$\left\|\cdot\right\|_{\rom{F}}$ for the matrix Frobenius norm, 
$\mathrm{diag}(a)$ for the diagonal matrix with the real vector $a\in\IR^n$ on its diagonal, and 
$\underline{\lambda}\left(A\right)$ and $\overline{\lambda}\left(A\right)$ respectively for the minimum and the maximum eigenvalues of the real matrix $A\in\IR^{n\times n}$. 

\section{Problem Formulation and Preliminaries}\label{problem_formulation}

In this paper, we consider the uncertain dynamical system represented in the state-space form given by
\bequ 
\ \ \ \dot{x}(t)=A x(t) + B \Lambda \bigl(u(t) +\delta(x(t))\bigl), \quad x(0)=x_0, \label{2.1}
\eequ 
where $x(t)\in\mathcal{D}\subseteq\IR^n$ is the measurable state  and $u(t)\in\IR^m$ is the control signal\footnote{Around an equilibrium point, a considerable number of uncertain dynamical systems including aerial and ground vehicles adhere to the state-space form given by (\ref{2.1}) either explicitly or approximately [\citen{yucelenMRAC}, Section 2].}. 
Here, $A\in\IR^{n \times n}$ is the known system matrix and $B\in\IR^{n \times m}$ is the known full column rank control matrix such that the pair $(A,B)$ is controllable. 
Furthermore, $\Lambda=\rom{diag}(\lambda)\in\IR_+^{m \times m}$, $\lambda=[\lambda_1,\ldots,\lambda_m]\mT$, is an unknown control effectiveness matrix and $\delta(x(t)):\mathcal{D}\rightarrow\IR^m$ is a parametric ($\mathcal{D}=\IR^n$ in this case) or nonparametric ($\mathcal{D}\subset\IR^n$ in this case with $\mathcal{D}$ being a compact set) uncertainty that is composed of locally Lipschitz functions. 

Next, let $u_\rom{n}(t)\in\IR^m$ be a nominal control signal that satisfies
\bequ 
    u_\rom{n}(t) = -K_1 x(t)+K_2 r(t), \label{2.2}
\eequ 
where $K_1\in\IR^{m \times n}$ is a feedback gain matrix such that $A-BK_1$ is Hurwitz, $K_2\in\IR^{m \times p}$ is a feedforward gain matrix, and $r(t)\in\IR^p$ is a given uniformly continuous and bounded reference signal\footnote{Without loss of generality, we use a static nominal control signal as given by (\ref{2.2}). 
If preferred, a dynamic nominal control signal can be employed (see [\citen{yucelenMRAC}, Section 3-B] for an example), where the results presented in this paper only experience minor modifications in this case.}. 
The purpose of (\ref{2.2}) is to define a nominal (i.e., ideal) closed-loop system behavior given by
\bequ
    \ \ \ \ \ \dot{x}_\rom{n}(t)=A_\rom{n}x_\rom{n}(t)+B_\rom{n}r(t), \quad x_\rom{n}(0)=x_{\rom{n}{0}}, \label{2.3}
\eequ 
in the absence of uncertainties (i.e., $\delta(x(t)) \equiv 0$ and $\Lambda \equiv I$) and $u(t)\equiv u_\rom{n}(t)$, where $A_\rom{n}\triangleq A-BK_1$ and $B_\rom{n}\triangleq B K_2$. 

\vspace{0.1cm}

To mitigate the effects of uncertainties, let $u_\rom{f}(t)\in\IR^m$ be the fixed-gain control signal and $u_\rom{a}(t)\in\IR^m$ be the adaptive learning signal. 
We can now define the control signal as
\bequ 
    u(t)=u_\rom{n}(t)+u_\rom{f}(t)+u_\rom{a}(t), \label{2.4}
\eequ 
with $u_\rom{f}(t)$ and $u_\rom{a}(t)$ being defined in the next sections. 
Using (\ref{2.4}) with (\ref{2.2}) in (\ref{2.1}), one can write
\bequ 
    \dot{x}(t)&=&A_\rom{n}x(t)+B_\rom{n}r(t)+B\Lambda\bigl(u_\rom{f}(t)+u_\rom{a}(t)\nonumber\\&&+\pi(x(t),u_\rom{n}(t))   \bigl), \label{2.5}
\eequ 
where 
\bequ \pi(x(t),u_\rom{n}(t))\triangleq \delta(x(t))+(I-\Lambda^{-1})u_\rom{n}(t) \label{total:unc}
\eequ 
is the total uncertainty. 

The problem studied in this paper is to synergistically integrate fixed-gain control signal $u_\rom{f}(t)$ and adaptive learning signal $u_\rom{a}(t)$ (i.e., symbiotic control) to mitigate the total uncertainty $\pi(x(t),u_\rom{n}(t))$ in a more predictable manner as compared to adaptive learning alone and without requiring any knowledge on this uncertainty. 
Section \ref{parametric_uncertainty} (respectively, Section \ref{nonparametric_uncertainty}) addresses this problem when $\delta(x(t))$ in $\pi(x(t),u_\rom{n}(t))$ is parametric (respectively, nonparametric), where we are now ready to cover important preliminaries in the next subsections of this section.  


\subsection{Preliminaries on Fixed-Gain Control}\label{fixed_gain_control} 

For the preliminaries presented in this subsection, let $u_\rom{a}(t)\equiv 0$ and consider the fixed-gain control signal 
\bequ
    u_\rom{f}(t)&\hspace{-0.25cm}=\hspace{-0.25cm}&-\alpha B_\rom{i}(x(t)\hspace{-0.05cm}-\hspace{-0.05cm}x_0)\hspace{-0.05cm}+\hspace{-0.05cm}\alpha B_\rom{i}\hspace{-0.05cm}\int_0^t \hspace{-0.05cm}\bigl(A_\rom{n}x(s)\hspace{-0.05cm}+\hspace{-0.05cm}B_\rom{n}r(s)\bigl)\rom{d}s, \nonumber\\ \label{2.6}
\eequ 
where $\alpha\in\IR_+$ is the fixed-gain control parameter and $B_\rom{i}\triangleq (B\mT B)^{-1}B\mT$. 
Note that since $B$ has full column rank, the inverse of $B\mT B$ exists. 
For instances where $x_0$ is not perfectly known, note also that its estimate can be used instead and this change only affects the initial transient closed-loop system behavior.  
We now present a key lemma. 

\vspace{0.1cm}

\textbf{Lemma 1.} The fixed-gain control signal given by (\ref{2.6}) is equivalent to
\vspace{0cm}
\bequ 
    \dot{u}_\rom{f}(t)&\hspace{-0.25cm}=\hspace{-0.25cm}&\hspace{-0.05cm}-\alpha \Lambda \bigl( u_\rom{f}(t)\hspace{-0.05cm}+\hspace{-0.05cm}\pi(x(t),u_\rom{n}(t))\bigl), \ \ u_\rom{f}(0)=0.  \label{2.7}
\eequ 

\vspace{0.1cm}

\textit{Proof.} Multiplying both sides of (\ref{2.5}) with $B_\rom{i}$ yields
\vspace{0cm}
\bequ
\Lambda\bigl(u_\rom{f}(t)\hspace{-0.05cm}+\hspace{-0.05cm}\pi(x(t),u_\rom{n}(t))   \bigl) \hspace{0.025cm}=\hspace{-0.075cm} B_\rom{i}\bigl(\dot{x}(t)\hspace{-0.075cm}-\hspace{-0.075cm}A_\rom{n}x(t)\hspace{-0.075cm}-\hspace{-0.075cm}B_\rom{n}r(t)\bigl),\hspace{-0.05cm} \label{lemma1.1}
\eequ 
where using (\ref{lemma1.1}) in (\ref{2.7}) gives
\bequ 
    \dot{u}_\rom{f}(t)&\hspace{-0.25cm}=\hspace{-0.25cm}&\hspace{-0.05cm}-\alpha B_\rom{i}\bigl(\dot{x}(t)-A_\rom{n}x(t)-B_\rom{n}r(t)\bigl).  \label{lemma1.2}
\eequ 
Finally, taking the integral of (\ref{lemma1.2}) results in (\ref{2.6}).  \hfill $\blacksquare$

\vspace{0.1cm}

While the fixed-gain control signal given by (\ref{2.6}) is implementable, its equivalent form given by (\ref{2.7}) is not. 
Because, the matrix $\Lambda$ and the term $\pi(x(t),u_\rom{n}(t))$ are unknown in (\ref{2.7}). 
While this is the case, (\ref{2.7}) is needed for an important result presented in the next proposition. 

\vspace{0.1cm}

\textbf{Proposition 1.} If $\alpha$ in (\ref{2.6}) is sufficiently large, then the solution to (\ref{2.5}) approximately behaves as the solution to the nominal (i.e., ideal) closed-loop system given by (\ref{2.3}). 

\vspace{0.1cm}

\textit{Proof.} Recall that (\ref{2.6}) is equivalent to (\ref{2.7}) by Lemma 1. 
Now, let $\varepsilon \triangleq \alpha^{-1}$ and rewrite (\ref{2.7}) as 
\vspace{-0.075cm}
\bequ 
    \varepsilon\dot{u}_\rom{f}(t)&\hspace{-0.25cm}=\hspace{-0.25cm}&\hspace{-0.05cm}- \Lambda \bigl( u_\rom{f}(t)\hspace{-0.05cm}+\hspace{-0.05cm}\pi(x(t),u_\rom{n}(t))\bigl), \label{2.11}
\eequ 
where (\ref{2.5}) and (\ref{2.11}) together represents a singularly perturbed dynamical system. 
In this case, a sufficiently large $\alpha$ implies that $\varepsilon\equiv 0$ in (\ref{2.11}), which results in the unique root 
\vspace{-0.075cm}
\bequ 
u_\rom{f}(t)=-\pi(x(t),u_\rom{n}(t)). \label{theorem1.1}
\eequ 
Using (\ref{theorem1.1}) in (\ref{2.5}), we arrive to the reduced-order dynamical system given by
\vspace{-0.075cm}
\bequ
    \ \ \ \ \ \dot{x}_\rom{r}(t)=A_\rom{n}x_\rom{r}(t)+B_\rom{n}r(t), \quad x_\rom{r}(0)=x_0, \label{2.10}
\eequ 
which has an exponentially stable equilibrium point. 
One can also write the boundary layer dynamical system as  
\vspace{-0.075cm}
\bequ 
\rom{d}u_\rom{fb}(\tau)/\rom{d}\tau=-\Lambda u_\rom{fb}(\tau), \quad u_\rom{fb}(0)=0, \quad t=\varepsilon \tau, 
\eequ 
which also has an exponentially stable equilibrium point since the unknown matrix $-\Lambda$ is Hurwitz by definition. 
All conditions of Tikhonov's theorem are now met [\citen{khalil}, Theorem 11.1]. 
Therefore, the solution to (\ref{2.5}) can be approximated using the solution to the reduced-order dynamical system given by (\ref{2.10}). 
Since the solution to (\ref{2.10}) is identical to the solution to (\ref{2.3}), the proof is now complete. \hfill $\blacksquare$

\vspace{0.1cm}

\textbf{Remark 1.} If the fixed-gain control parameter $\alpha$ is sufficiently large, then it is evident from Proposition 1 that the solution to the uncertain dynamical system given by (\ref{2.5}) approximates the solution to the nominal (i.e., ideal) closed-loop system given by (\ref{2.3}). 
This is a strong result that holds for both parametric or nonparametric uncertainty cases and without an adaptive learning signal. 
However, it is not possible to know in practice how large $\alpha$ needs to be. 
Furthermore, analyzing closed-loop system stability on how $\alpha$ needs to be properly chosen without an adaptive learning signal requires a specific uncertainty structure and an upper bound on this uncertainty, where such analysis can be conservative as well (see [\citen{de2016new}] for an examplary study involving small gain theorem). 
To avoid all these, we employ a form of the fixed-gain control signal given by (\ref{2.6}) in the following sections with an adaptive learning signal (i.e., symbiotic control). 
Importantly, we harness the presence of this fixed-gain control signal for achieving a more predictable closed-loop system behavior as $\alpha$ increases. 

\vspace{0.1cm}


\subsection{Preliminaries on Adaptive Learning}\label{adaptive_learning} 

For the preliminaries presented in this subsection, let $u_\rom{f}(t)\equiv 0$. 
First, consider that $\delta(x(t))$ is a parametric uncertainty such that
\vspace{-0.05cm}
\bequ
\delta(x(t))=W\mT_\delta \sigma_\delta(x(t)), \quad x(t)\in\IR^n, \label{parameterized:unc:here}
\eequ 
holds, where $W_\delta\in\IR^{s\times m}$ is an unknown weight and $\sigma_\delta(x(t)):\IR^n \rightarrow \IR^s$ is a known basis function. 
In this case, the total uncertainty given by (\ref{total:unc}) can be represented as
\bequ \pi(x(t),u_\rom{n}(t))\triangleq W\mT \sigma(x(t),u_\rom{n}(t)), \label{total:unc:2}
\eequ 
where $W\triangleq [W\mT_\delta,(I-\Lambda^{-1})]\mT\in\IR^{(s+m) \times m}$ is unknown and $\sigma(x(t),u_\rom{n}(t))\triangleq [\sigma\mT_\delta(x(t)),u\mT_\rom{n}(t)]\mT:\IR^n \times \IR^m \rightarrow \IR^{s+m}$ is known by definition. 
In addition, let the adaptive learning signal be
\vspace{-0.05cm}
\bequ 
u_\rom{a}(t)=-\hat{W}\mT(t) \sigma(x(t),u_\rom{n}(t)), \label{prm:1}
\eequ 
where $\hat{W}(t)\in\IR^{(s+m) \times m}$ is an estimate of $W$ that is learned through the parameter adjustment mechanism given by 
\vspace{-0.05cm}
\bequ 
\dot{\hat{W}}(t)&\hspace{-0.275cm}=\hspace{-0.275cm}&\beta \sigma(x(t),u_\rom{n}(t)) \bigl(x(t)\hspace{-0.03cm}-\hspace{-0.03cm}x_\rom{n}(t)\bigl)\mT PB, \ \hat{W}(0)=\hat{W}_0, \nonumber\\&& \label{prm:2}
\eequ 
with $\beta\in\IR_+$ being the adaptive learning parameter, $x_\rom{n}(t)$ satisfying (\ref{2.3}), and $P\in\IR_+^{n\times n}$ being the unique solution to the Lyapunov function 
\vspace{-0.05cm}
\bequ 
0=A\mT_\rom{n}P+PA_\rom{n}+R \label{L:E}
\eequ 
for a given $R\in\IR_+^{n\times n}$. 
Note that if $x_0$ is not perfectly known, its estimate can be used instead to initialize (\ref{2.3}) and this change only affects the initial transient closed-loop system behavior. 
We are now ready for the next proposition. 

\vspace{0.1cm}

\textbf{Proposition 2\footnote{See [\citen{yucelenMRAC}, Section 3-C] for the proof of this proposition.}.} Consider the dynamical system given by (\ref{2.5}) with the parametric uncertainty given by (\ref{parameterized:unc:here}), where $u_\rom{f}(t)\equiv 0$. 
Consider also the adaptive learning signal given by (\ref{prm:1}), (\ref{prm:2}), and (\ref{2.3}). 
The trajectories of the closed-loop system are then bounded and $\lim_{t\rightarrow \infty}\bigl(x(t)-x_\rom{n}(t)\bigl)=0$.

\vspace{0.1cm}

Next, consider that $\delta(x(t))$ is a nonparametric uncertainty such that
\vspace{-0.05cm}
\bequ
\delta(x(t))=W\mT_\delta \sigma_\delta(x(t))+\epsilon(x(t)), \quad x(t)\in\mathcal{D}, \label{non:prm:unc}
\eequ 
holds, where $W_\delta\in\IR^{s\times m}$ is an unknown weight, $\sigma_\delta(x(t)):\mathcal{D} \rightarrow \IR^s$ is a known basis function that contains a unity bias and $s_\rom{f} \triangleq s-1$ radial basis functions, and $\epsilon(x(t))$ is a bounded approximation error\footnote{By the Weierstrass approximation theorem \citen{hornik1990universal}, this feedforward radial basis function neural network guarantees that $\epsilon(x(t))\rightarrow 0$ as $s_\rom{f} \rightarrow \infty$.}. 
In this case, the total uncertainty (\ref{total:unc}) can be represented as
\vspace{0cm}
\bequ \pi(x(t),u_\rom{n}(t))\triangleq W\mT \sigma(x(t),u_\rom{n}(t))+\epsilon(x(t)), \label{total:unc:3}
\eequ 
where $W$ and $\sigma(x(t),u_\rom{n}(t))$ have the same forms given after (\ref{total:unc:2}). 
In addition, let the adaptive learning signal be given by (\ref{prm:1}), 
where $\hat{W}(t)$ is an estimate of $W$ that is learned through the parameter adjustment mechanism given by 
\vspace{0cm}
\bequ 
\dot{\hat{W}}(t)&\hspace{-0.275cm}=\hspace{-0.275cm}&\beta_1 \sigma(x(t),u_\rom{n}(t)) \bigl(x(t)\hspace{-0.03cm}-\hspace{-0.03cm}x_\rom{n}(t)\bigl)\mT PB-\beta_2\hat{W}(t), \nonumber\\&& \hspace{4.25cm} \hat{W}(0)=\hat{W}_0, \label{prm:10}
\eequ 
with $\beta_1\in\IR_+$ being the adaptive learning parameter, $\beta_2\in\IR_+$ being the leakage parameter, $x_\rom{n}(t)$ satisfying (\ref{2.3}), and $P\in\IR_+^{n\times n}$ being the unique solution to the Lyapunov equation (\ref{L:E}). 
We are now ready for the next proposition. 

\vspace{0.1cm}

\textbf{Proposition 3\footnote{From [\citen{yucelenMRAC}, Section 3-E], the proof of this proposition follows.}.} Consider the dynamical system given by (\ref{2.5}) with the nonparametric uncertainty given by (\ref{non:prm:unc}), where $u_\rom{f}(t)\equiv 0$. 
Consider also the adaptive learning signal given by (\ref{prm:1}), (\ref{prm:10}), and (\ref{2.3}). 
The trajectories of the closed-loop system are then bounded. 

\vspace{0.1cm}

\textbf{Remark 2.} A form of direct adaptive control method given in Propositions 2 and 3 is used in the next sections for constructing the adaptive learning signal $u_\rom{a}(t)$. 
Because, direct adaptive control does not require persistency of excitation for mitigating the effects of uncertainties. 
If preferred, one can also explore other adaptive learning methods to use together with fixed-gain control. 
However, it is worth noting that direct adaptive control, like other adaptive learning methods, can exhibit less predictable, poor closed-loop system behavior due to their nonlinear parameter adjustment mechanism and in the presence of high neural network approximation errors (i.e., $\epsilon(x(t))$ in (\ref{non:prm:unc})) and high leakage parameter (i.e., $\beta_2$ in (\ref{prm:10})). 
We address this issue by synergistically integrating the fixed-gain control signal with the adaptive learning signal (i.e., symbiotic control) in the next sections.

\vspace{0.1cm}

\textbf{Remark 3.} The following two observations about Proposition 3 are now given. 
First, it holds when $x(t)$ stays in $\mathcal{D}$. 
To enforce $x(t)$ to stay in $\mathcal{D}$ without necessarily making $\mathcal{D}$ arbitrarily large, one can use set-theoretic direct adaptive control method [\citen{arabi2019neuroadaptive}]. 
Second, Proposition 3 holds when one uses a projection operator in (\ref{prm:10}) instead of the leakage term (i.e., $-\beta_2\hat{W}(t)$). 
We prefer not to use a projection operator in order not to make any assumptions on the bounds of $W$. 

\vspace{0.1cm}


\subsection{Preliminaries on Composite Function Construction}\label{function_construction} 

\begin{figure}[t!] \vspace{-0.0cm}
\hspace{0.3cm}\includegraphics[width=0.44\textwidth]{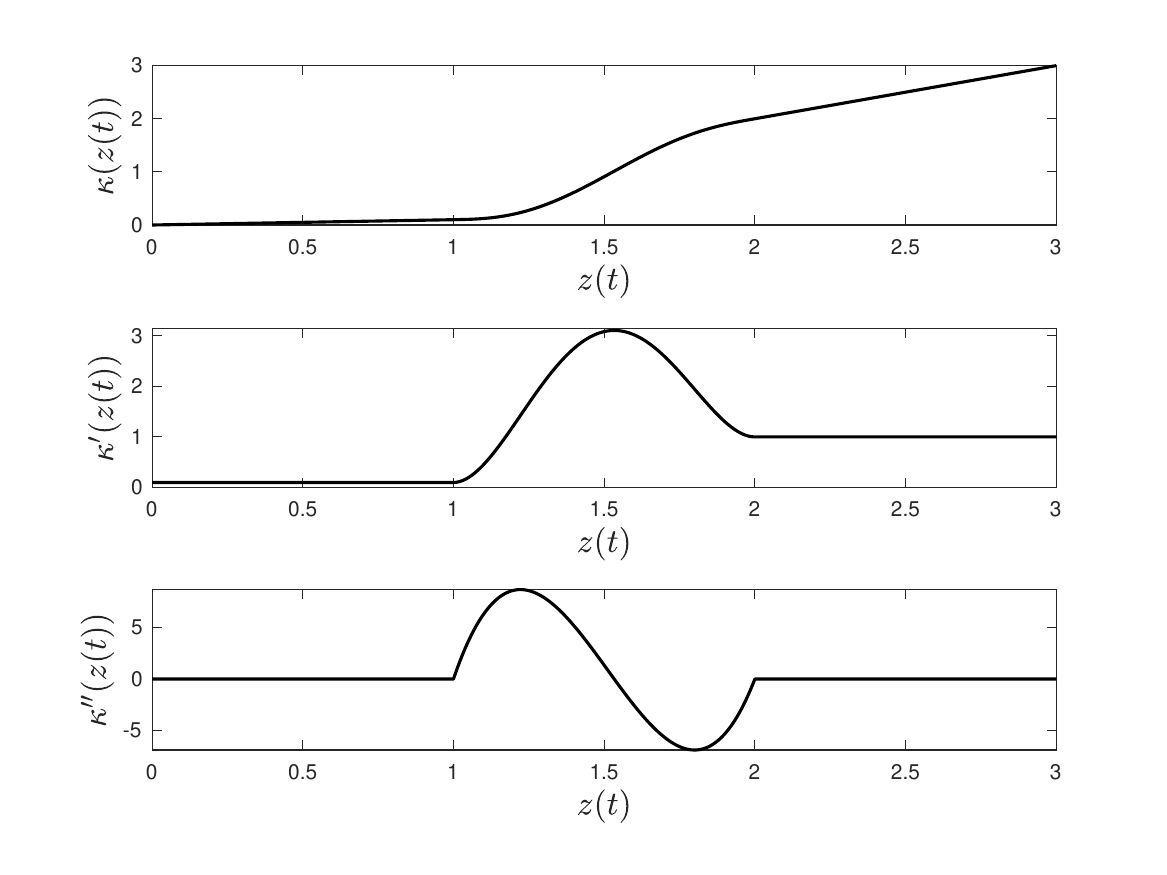} \vspace{-0.6cm}
    \caption{Composite function $\kappa(z(t))$ given by (\ref{fc:1}) for $a=1$, $b=2$, and $\rho=0.1$, where $\kappa'(z(t)) \triangleq \rom{d} \kappa(z(t))/\rom{d}z(t)$ and $\kappa''(z(t)) \triangleq \rom{d}^2 \kappa(z(t))/\rom{d}z^2(t)$. } \vspace{-0.4cm}
    \label{composite:function} 
\end{figure}

We construct a composite function given by
\vspace{0cm}
\begin{equation}
  \kappa(z(t)) =
    \begin{cases}
      \rho z(t), & \text{$z(t)\in[0,a]$,}\\
      \kappa_\rom{c}(z(t)), & \text{$z(t)\in(a,b)$,}\\
      z(t), & \text{$z(t)\in[b,\infty)$,}
    \end{cases}       \label{fc:1}
\end{equation}
where $z(t)\in\IR_+$, $\rho\in[0,1)$, $a\in\IR_+$, and $b\in\IR_+$ such that $a<b$. 
In (\ref{fc:1}), $\kappa_\rom{c}(z(t)):\IR_+\rightarrow\IR_+$ is a transition function that is chosen as 
\vspace{0cm}
\bequ 
\kappa_\rom{c}(z(t)) &=& \sum_{k=0}^l \psi_k z^k(t). \label{fc:2}
\eequ 
Letting $\kappa'_\rom{c}(z(t)) \triangleq \rom{d} \kappa_\rom{c}(z(t))/\rom{d}z(t)=\sum_{k=1}^l k\psi_k z^{k-1}(t)$ and $\kappa''_\rom{c}(z(t)) \triangleq \rom{d}^2 \kappa_\rom{c}(z(t))/\rom{d}z^2(t)=\sum_{k=2}^l k(k-1)\psi_k z^{k-2}(t)$, the conditions $\kappa_\rom{c}(a)=\rho a$, $\kappa'_\rom{c}(a)=\rho$, $\kappa_\rom{c}''(a)=0$, $\kappa_\rom{c}(b)= b$, $\kappa_\rom{c}'(b)=1$, and $\kappa_\rom{c}''(b)=0$ are needed to hold for (\ref{fc:1}) to be twice continuously differentiable. 
To this end, using $\kappa_\rom{c}(z(t))$, $\kappa'_\rom{c}(z(t))$, and $\kappa''_\rom{c}(z(t))$ in these conditions for $l=5$, we arrive
\bequ 
\ \ \ {\footnotesize{\begin{bmatrix}
1 & a & a^2 & a^3 & a^4 & a^5\\
0 & 1 & 2 a & 3 a^2 & 4 a^3 & 5 a^4\\
0 & 0 & 2 & 6 a & 12 a^2 & 20 a^3\\
1 & b & b^2 & b^3 & b^4 & b^5\\
0 & 1 & 2 b & 3 b^2 & 4 b^3 & 5 b^4\\
0 & 0 & 2 & 6 b & 12 b^2 & 20 b^3
\end{bmatrix} 
\begin{bmatrix}
\psi_0 \\
\psi_1 \\
\psi_2 \\
\psi_3 \\
\psi_4 \\
\psi_5 
\end{bmatrix}=\begin{bmatrix}
\rho a \\
\rho \\
0 \\
b \\
1 \\
0
\end{bmatrix},}}  \label{fc:5}
\eequ
where solution $\psi_i$, $i=0,\ldots,5$, to (\ref{fc:5}) gives the parameters to implement the transition function (\ref{fc:2}), where this solution should always exist [\citen{muse2011method}]. 
Figure \ref{composite:function} illustrates the composite function $\kappa(z(t))$ given by (\ref{fc:1}) for $a=1$, $b=2$, and $\rho=0.1$. 

\vspace{0.1cm}

\textbf{Remark 4.} We draw attention to two points. 
First, we require that $\kappa(z(t))\in\overline{\IR}_+$ and $\kappa'(z(t)) \in\overline{\IR}_+$ when $\rho=0$, and $\kappa(z(t))\in{\IR}_+$ and $\kappa'(z(t)) \in{\IR}_+$ otherwise\footnote{The composite function $\kappa(z(t))$ is not unique and one can explore its other versions with the results presented in the next sections.}. 
Second, the gradient $\kappa'(z(t))$ of the composite function (\ref{fc:1}) can be made sufficiently small over $z(t)\in[0,a]$ by selecting $\rho$ sufficiently small (e.g., $\kappa'(z(t))=0.1$ for $z(t)\in[0,1]$ in Figure \ref{composite:function}). 
In the next sections, this gradient is used to decrease the effect of the nonlinear terms resulting from the standard adaptive learning method outlined in Section \ref{adaptive_learning} when the trajectories of the uncertain dynamical system (\ref{total:unc}) and the trajectories of the nominal (i.e., ideal) closed-loop system (\ref{2.3}) remain close to each other. 

\section{Symbiotic Control of Dynamical Systems with Parametric Uncertainty}\label{parametric_uncertainty}

Consider the uncertain dynamical system given by (\ref{2.5}) subject to (\ref{total:unc}). 
In this section, we focus on the parametric uncertainty case such that (\ref{parameterized:unc:here}) holds over $x(t)\in\IR^n$, where the total uncertainty is represented as (\ref{total:unc:2}). 


\subsection{Control Architecture} \label{control_architecture_1}

To begin with, consider a form of the fixed-gain control signal given by
\vspace{-0.2cm}
\bequ
    u_\rom{f}(t)&\hspace{-0.25cm}=\hspace{-0.25cm}&-\alpha B_\rom{i}(x(t)\hspace{-0.05cm}-\hspace{-0.05cm}x_0)\hspace{-0.05cm}+\hspace{-0.05cm}\alpha B_\rom{i}\hspace{-0.05cm}\int_0^t \hspace{-0.05cm}\bigl(A_\rom{n}x(s)\hspace{-0.05cm}+\hspace{-0.05cm}B_\rom{n}r(s)\bigl)\rom{d}s \nonumber\\ &&+\int_0^t u_\rom{g}(s) \rom{d}s, \label{ca1.1}
\eequ 
where $\alpha\in\IR_+$ is the fixed-gain control parameter and $B_\rom{i}\triangleq (B\mT B)^{-1}B\mT$. 
In (\ref{ca1.1}), $u_\rom{g}(t)\in\IR^m$ satisfies
\vspace{0cm}
\bequ
    u_\rom{g}(t) &\hspace{-0.25cm}=\hspace{-0.25cm}& -\beta_1 \beta_2^{-1} \kappa'\bigl(e\mT(t)Pe(t)\bigl)\hat{\Lambda}(t)B\mT P e(t), \ \ \ \ \label{ca1.2}
\eequ 
where $e(t)\triangleq x(t)-x_\rom{n}(t)\in\IR^n$ is the error signal, 
$\kappa'\bigl(e\mT(t)Pe(t)\bigl) \triangleq \rom{d} \kappa\bigl(e\mT(t)Pe(t)\bigl)/\rom{d}\bigl(e\mT(t)Pe(t)\bigl)$ with $\kappa\bigl(e\mT(t)Pe(t)\bigl)$ satisfying (\ref{fc:1}), and $P\in\IR_+^{n\times n}$ is the unique solution to the Lyapunov function (\ref{L:E}) for a given $R\in\IR_+^{n\times n}$ ($\beta_1\in\IR_+$ and  $\beta_2\in\IR_+$ are defined below). 
In (\ref{ca1.2}), $\hat{\Lambda}(t)\in\IR^{m \times m}$ is an estimate of $\Lambda$ that is learned through the parameter adjustment mechanism 
\bequ
   \hspace{-0.25cm} \dot{\hat{\Lambda}}(t) &\hspace{-0.25cm}=\hspace{-0.25cm}& \gamma \kappa'\bigl(e\mT(t)Pe(t)\bigl) B\mT P e(t) u\mT_\rom{f}(t), \ \hat{\Lambda}(0)=\hat{\Lambda}_0, \ \ \ \label{ca1.3}
\eequ 
with $\gamma\in\IR_+$ being the adaptive learning parameter. 
Next, consider the adaptive learning signal given by (\ref{prm:1}), where $\hat{W}(t)\in\IR^{(s+m) \times m}$ is an estimate of $W$ that is learned through the parameter adjustment mechanism  
\bequ 
\dot{\hat{W}}(t)&\hspace{-0.275cm}=\hspace{-0.275cm}&\beta_1 \kappa'\bigl(e\mT(t)Pe(t)\bigl) \sigma(x(t),u_\rom{n}(t)) e\mT(t) PB, \nonumber\\
&&-\beta_2 \alpha\sigma(x(t),u_\rom{n}(t)) u\mT_\rom{f}(t), \ \hat{W}(0)=\hat{W}_0,  \label{ca1.4}
\eequ 
with $\beta_1\in\IR_+$ and  $\beta_2\in\IR_+$ being the adaptive learning parameters. 

\vspace{0.1cm}

\textbf{Remark 5.} The proposed symbiotic control architecture for dynamical systems with parametric uncertainty is given above\footnote{To implement this architecture, one primarily needs to choose $\alpha$ for (\ref{ca1.1}), $\gamma$ for (\ref{ca1.3}), $\beta_1$ and  $\beta_2$ for (\ref{ca1.4}), and $\rho$ for (\ref{fc:1}) (one can set $R=I$ for (\ref{L:E}), $\hat{\Lambda}_0=0$ for (\ref{ca1.3}), and $\hat{W}_0=0$ for (\ref{ca1.4})). 
{\textcolor{black}{We refer to Section \ref{example_1} for the selection of these parameters}}.}. 
Observe that the proposed fixed-gain control signal (\ref{ca1.1}) is a version of the original fixed-gain control signal (\ref{2.6}) with an added integral of (\ref{ca1.2}) containing the adaptive parameter $\hat{\Lambda}(t)$. 
Likewise, (\ref{ca1.4}) is a version of (\ref{prm:2}) with an added term containing the fixed-gain control signal $u_\rom{f}(t)$. 
In other words, the fixed-gain control and adaptive learning architectures interact with each other to mitigate the effects of uncertainties in a more predictable manner without requiring any knowledge of such uncertainties. 

\vspace{0.1cm}


\subsection{System-Theoretical Analysis} \label{analysis_1}

We begin with a key lemma. 

\vspace{0.1cm}

\textbf{Lemma 2.} The form of the fixed-gain control signal given by (\ref{ca1.1}) is equivalent to
\vspace{0cm}
\bequ 
    \dot{u}_\rom{f}(t)&\hspace{-0.25cm}=\hspace{-0.25cm}&\hspace{-0.05cm}-\alpha \Lambda \bigl( u_\rom{f}(t)\hspace{-0.05cm}+\hspace{-0.05cm}u_\rom{a}(t)\hspace{-0.05cm}+\hspace{-0.05cm}\pi(x(t),u_\rom{n}(t))\bigl)+u_\rom{g}(t), \nonumber\\&&\hspace{4.3cm} \ u_\rom{f}(0)=0.  \label{a.1}
\eequ 

\vspace{0.1cm}

\textit{Proof.} Multiplying both sides of (\ref{2.5}) with $B_\rom{i}$ yields
\vspace{0cm}
\bequ
&&\hspace{-1cm}\Lambda\bigl(u_\rom{f}(t)+u_\rom{a}(t)+\pi(x(t),u_\rom{n}(t))   \bigl) \nonumber\\
&&\hspace{1cm}= B_\rom{i}\bigl(\dot{x}(t)-A_\rom{n}x(t)-B_\rom{n}r(t)\bigl), \label{lemma1.1n}
\eequ 
where using (\ref{lemma1.1n}) in (\ref{a.1}) gives
\bequ 
    \dot{u}_\rom{f}(t)&\hspace{-0.25cm}=\hspace{-0.25cm}&\hspace{-0.05cm}-\alpha B_\rom{i}\bigl(\dot{x}(t)-A_\rom{n}x(t)-B_\rom{n}r(t)\bigl)+u_\rom{g}(t). \ \ \ \label{a.1b}
\eequ 
Finally, taking the integral of (\ref{a.1b}) results in (\ref{ca1.1}). \hfill $\blacksquare$

\vspace{0.1cm} 

While the form of the fixed-gain control signal given by (\ref{ca1.1}) is implementable, its equivalent form given by (\ref{a.1}) is not since the matrix $\Lambda$ and the term $\pi(x(t),u_\rom{n}(t))$ are unknown. 
This equivalent form is only needed for the system-theoretical analysis presented in this subsection. 
We are now ready to present an important theorem. 

\vspace{0.1cm}

\textbf{Theorem 1.} If $\alpha$ in (\ref{ca1.1}) is sufficiently large, then the solution to (\ref{2.5}) approximately behaves as the solution to the nominal (i.e., ideal) closed-loop system given by (\ref{2.3}). 

\vspace{0.1cm}

\textit{Proof.} Let $\zeta(t)\triangleq -u_\rom{a}(t)-\pi(x(t),u_\rom{n}(t))$, where (\ref{a.1}) can be rewritten in the Laplace domain as $u_\rom{f}(s)=\mathcal{G}_1(s)\zeta(s)+\mathcal{G}_2(s) u_\rom{g}(s)$, where $\mathcal{G}_1(s)\triangleq (sI+\alpha \Lambda)^{-1}(\alpha \Lambda)$, $\mathcal{G}_2(s)\triangleq (sI+\alpha \Lambda)^{-1}$, and $s$ is the Laplace variable\footnote{As opposed to the proof of Proposition 1, we may not be able to use singular perturbation theory here since the origin of the reduced-order dynamical system may not be exponentially stable due to the presence of the adaptive learning architecture [\citen{khalil}, Example 9.13]. This is the reason why we do the current proof in the Laplace domain.}. 
As $\alpha$ becomes sufficiently large (i.e., $\alpha\rightarrow \infty$), the magnitude and the phase of $\mathcal{G}_1(s)$ (respectively, $\mathcal{G}_2(s)$) respectively approach $0$ decibels and $0$ degrees (respectively, $-\infty$ decibels and $0$ degrees). 
This implies that $u_\rom{f}(t)$ approaches $\zeta(t)$, and hence, the solution to (\ref{2.5}) approaches the solution to (\ref{2.3}) as $\alpha$ becomes sufficiently large. \hfill $\blacksquare$

\vspace{0.1cm}

Theorem 1 shows that the closed-loop system behavior becomes more predictable as $\alpha$ increases, where a discussion similar to the one given in Remark 1 also applies to this theorem. 
Next, using (\ref{total:unc:2}) and (\ref{prm:1}) in (\ref{a.1}), we arrive
\bequ
    \dot{u}_\rom{f}(t)  &\hspace{-0.25cm}=\hspace{-0.25cm}& -\alpha \Lambda\bigl(u_\rom{f}(t)-\tilde{W}\mT(t)\sigma(x(t),u_\rom{n}(t))\bigl)+u_\rom{g}(t), \ \ \ \ \ \label{a.2}
\eequ 
where $\tilde{W}(t)\triangleq \hat{W}(t)-W \in\IR^{(s+m) \times m}$. 
Furthermore, using (\ref{2.5}) and (\ref{2.3}), the time derivative of the error signal is written in the form given by
\bequ 
    \dot{e}(t) &\hspace{-0.25cm}=\hspace{-0.25cm}& A_\rom{n}e(t)+B\Lambda\bigl(u_\rom{f}(t)+u_\rom{a}(t)+\pi(x(t),u_\rom{n}(t))   \bigl), \nonumber\\ &&\hspace{4.9cm} e(0)=0, \label{a.3}
\eequ 
Once again, using (\ref{total:unc:2}) and (\ref{prm:1}) in (\ref{a.3}), we arrive
\bequ 
    \dot{e}(t)&\hspace{-0.25cm}=\hspace{-0.25cm}&A_\rom{n}e(t)+B\Lambda\bigl(u_\rom{f}(t)-\tilde{W}\mT(t)\sigma(x(t),u_\rom{n}(t))   \bigl). \ \ \ \ \ \label{a.4}
\eequ 
We are now ready to present our first main result. 
For this purpose, we also define $\tilde{\Lambda}(t)\triangleq \hat{\Lambda}(t)-\Lambda\in\IR^{m \times m}$. 

\vspace{0.1cm}

\textbf{Theorem 2.} Consider the dynamical system given by (\ref{2.5}) with the parametric uncertainty given by (\ref{parameterized:unc:here}). 
In addition, consider the form of the fixed-gain control signal given by (\ref{ca1.1}) with (\ref{ca1.2}) and (\ref{ca1.3}). 
Consider also the form of the adaptive learning signal given by (\ref{prm:1}), (\ref{ca1.4}), and (\ref{2.3}). 
The trajectories $\bigl(e(t),u_\rom{f}(t),\tilde{W}(t),\tilde{\Lambda}(t)\bigl)$ of the closed-loop system are then bounded and
\bequ
    \lim_{t\rightarrow \infty} \Bigl(\kappa'\bigl(e\mT(t)P e(t)\bigl)e(t), \hspace{0.03cm}u_\rom{f}(t)\Bigl)=\bigl(0, \hspace{0.03cm}0\bigl). \label{result:1}
\eequ

\vspace{0.1cm}

\textit{Proof.} Consider the energy function given by
\bequ
\mathcal{V}(\cdot) &\hspace{-0.25cm}=\hspace{-0.25cm}& \beta_1 \kappa \bigl(e\mT(t) P e(t)\bigl)+\beta_2 u_\rom{f}\mT(t)u_\rom{f}(t)\nonumber\\&&
+\rom{tr} \bigl(\tilde{W}(t)\Lambda^{\frac{1}{2}}\bigl)\mT  \bigl(\tilde{W}(t)\Lambda^{\frac{1}{2}}\bigl)\nonumber\\&&
+\beta_1\gamma^{-1}\rom{tr}\tilde{\Lambda}\mT (t) \tilde{\Lambda}(t). \label{st:01}
\eequ 
Note that $\mathcal{V}(0,0,0,0)=0$, $\mathcal{V}(e(t),u_\rom{f}(t),\tilde{W}(t),\tilde{\Lambda}(t))\in\IR_+$ when $\rho\neq 0$ in (\ref{fc:1}) and $\mathcal{V}(e(t),u_\rom{f}(t),\tilde{W}(t),\tilde{\Lambda}(t))\in\overline{\IR}_+$ otherwise, and $\mathcal{V}(e(t),u_\rom{f}(t),\tilde{W}(t),\tilde{\Lambda}(t))$ is radially unbounded. 
The time derivative of (\ref{st:01}) satisfies
\vspace{0cm}
\bequ 
    \dot{\mathcal{V}}(\cdot) &\hspace{-0.25cm}=\hspace{-0.25cm}& 
    2 \beta_1 \kappa'(\cdot) e\mT(t) P \bigl[A_\rom{n}e(t)\hspace{-0.05cm}+\hspace{-0.05cm}B\Lambda\bigl(u_\rom{f}(t)  
    \hspace{-0.05cm}-\hspace{-0.05cm}\tilde{W}\mT(t)\sigma(\cdot)   \bigl)\bigl] \nonumber\\&&
    +2\beta_2 u_\rom{f}\mT(t)\bigl[  -\alpha \Lambda\bigl(u_\rom{f}(t)-\tilde{W}\mT(t)\sigma(\cdot)\bigl)+u_\rom{g}(t) \bigl]\nonumber\\&&
    +2\rom{tr}\tilde{W}\mT(t)\bigl[ \beta_1 \kappa'(\cdot)\sigma(\cdot) e\mT(t) PB  
    -\beta_2 \alpha\sigma(\cdot)\nonumber\\&& \times u\mT_\rom{f}(t) \bigl]\Lambda
    +2\beta_1 \rom{tr}\tilde{\Lambda}\mT(t) \bigl[ \kappa'(\cdot) B\mT P e(t) u\mT_\rom{f}(t)    ] \nonumber\\
    &\hspace{-0.25cm}=\hspace{-0.25cm}&-\beta_1\kappa'(\cdot)e\mT(t)R e(t)+2\beta_1\kappa'(\cdot)e\mT(t)PB\Lambda u_\rom{f}(t) \nonumber\\
    && -2\alpha \beta_2 u_\rom{f}\mT(t)\Lambda u_\rom{f}(t)-2\beta_1 \kappa'(\cdot) u_\rom{f}\mT(t)\hat{\Lambda}(t)B\mT P e(t) \nonumber\\
    && +2\beta_1 \rom{tr}\tilde{\Lambda}\mT(t) \bigl[ \kappa'(\cdot) B\mT P e(t) u\mT_\rom{f}(t)    ] \nonumber\\
    &\hspace{-0.25cm}=\hspace{-0.25cm}& -\beta_1\kappa'(\cdot)e\mT(t)R e(t)-2\alpha \beta_2 u_\rom{f}\mT(t)\Lambda u_\rom{f}(t) \nonumber\\
    && -2\beta_1 \kappa'(\cdot) u_\rom{f}\mT(t)\tilde{\Lambda}(t)B\mT P e(t) \nonumber\\
    && +2\beta_1 \rom{tr}\tilde{\Lambda}\mT(t) \bigl[ \kappa'(\cdot) B\mT P e(t) u\mT_\rom{f}(t)    ] \nonumber\\
    &\hspace{-0.25cm}=\hspace{-0.25cm}& -\beta_1\kappa'(\cdot)e\mT(t)R e(t)-2\alpha \beta_2 u_\rom{f}\mT(t)\Lambda u_\rom{f}(t), \label{st:02}
\eequ 
which implies the boundedness of the closed-loop system trajectories $\bigl(e(t),u_\rom{f}(t),\tilde{W}(t),\tilde{\Lambda}(t)\bigl)$. 
From the LaSalle-Yoshizawa theorem [\citen{haddad2008nonlinear}, Theorem 4.7], (\ref{result:1}) now holds. \hfill $\blacksquare$

\vspace{0.1cm}

\section{Symbiotic Control of Dynamical Systems with Nonparametric Uncertainty}\label{nonparametric_uncertainty}

Consider the uncertain dynamical system given by (\ref{2.5}) subject to (\ref{total:unc}). 
In this section, we focus on the nonparametric uncertainty case such that (\ref{non:prm:unc}) holds over $x(t)\in\mathcal{D}$, where the total uncertainty is represented as (\ref{total:unc:3}). 


\subsection{Control Architecture} \label{control_architecture_2}

To begin with, consider a form of the fixed-gain control signal given by (\ref{ca1.1}), where $\alpha\in\IR_+$ is the fixed-gain control parameter and $B_\rom{i}\triangleq (B\mT B)^{-1}B\mT$. 
In (\ref{ca1.1}), $u_\rom{g}(t)\in\IR^m$ satisfies (\ref{ca1.2}), where $e(t)\triangleq x(t)-x_\rom{n}(t)\in\IR^n$ is the error signal, 
$\kappa'\bigl(e\mT(t)Pe(t)\bigl) \triangleq \rom{d} \kappa\bigl(e\mT(t)Pe(t)\bigl)/\rom{d}\bigl(e\mT(t)Pe(t)\bigl)$ with $\kappa\bigl(e\mT(t)Pe(t)\bigl)$ satisfying (\ref{fc:1}), and $P\in\IR_+^{n\times n}$ is the unique solution to the Lyapunov function (\ref{L:E}) for a given $R\in\IR_+^{n\times n}$ ($\beta_1\in\IR_+$ and  $\beta_2\in\IR_+$ are defined below). 
In (\ref{ca1.2}), $\hat{\Lambda}(t)\in\IR^{m \times m}$ is an estimate of $\Lambda$ that is learned through the parameter adjustment mechanism 
\bequ
   \hspace{-0.25cm} \dot{\hat{\Lambda}}(t) &\hspace{-0.25cm}=\hspace{-0.25cm}& \gamma_1 \kappa'\bigl(e\mT(t)Pe(t)\bigl) B\mT P e(t) u\mT_\rom{f}(t)-\gamma_2 \hat{\Lambda}(t), \nonumber\\ &&\hspace{4.4cm} \hat{\Lambda}(0)=\hat{\Lambda}_0,  \label{ca2.3}
\eequ 
with $\gamma_1\in\IR_+$ being the adaptive learning parameter and $\gamma_2\in\IR_+$ being the leakage parameter.  
Next, consider the adaptive learning signal given by (\ref{prm:1}), where $\hat{W}(t)\in\IR^{(s+m) \times m}$ is an estimate of $W$ that is learned through the parameter adjustment mechanism  
\bequ 
\dot{\hat{W}}(t)&\hspace{-0.275cm}=\hspace{-0.275cm}&\beta_1 \kappa'\bigl(e\mT(t)Pe(t)\bigl) \sigma(x(t),u_\rom{n}(t)) e\mT(t) PB \nonumber\\
&&-\beta_2 \alpha\sigma(x(t),u_\rom{n}(t)) u\mT_\rom{f}(t)-\beta_3 \hat{W}(t), \nonumber\\
&&\hspace{3.9cm} \hat{W}(0)=\hat{W}_0,  \label{ca2.4}
\eequ 
with $\beta_1\in\IR_+$ and  $\beta_2\in\IR_+$ being the adaptive learning parameters, and $\beta_3\in\IR_+$ being the leakage parameter. 
Note that similar version of Remark 5 also applies to the proposed symbiotic control architecture given above for dynamical systems with nonparametric uncertainty\footnote{To implement this architecture, one primarily needs to choose $\alpha$ for (\ref{ca1.1}); $\gamma_1$ and $\gamma_2$ for (\ref{ca2.3}); $\beta_1$, $\beta_2$, and $\beta_3$ for (\ref{ca2.4}); and $\rho$ for (\ref{fc:1}) (one can set $R=I$ for (\ref{L:E}), $\hat{\Lambda}_0=0$ for (\ref{ca2.3}), and $\hat{W}_0=0$ for (\ref{ca2.4})). 
{\textcolor{black}{We refer to Section \ref{example_2} for the selection of these parameters}}.}. 


\subsection{System-Theoretical Analysis} \label{analysis_2}

We begin with the note that Lemma 2 and Theorem 1 also hold for the nonparametric uncertainty case. 
Next, using (\ref{total:unc:3}) and (\ref{prm:1}) in (\ref{a.1}), we arrive
\bequ
    \dot{u}_\rom{f}(t)  &\hspace{-0.25cm}=\hspace{-0.25cm}& -\alpha \Lambda\bigl(u_\rom{f}(t)-\tilde{W}\mT(t)\sigma(x(t),u_\rom{n}(t))+\epsilon(x(t))\bigl)\nonumber\\&&+u_\rom{g}(t), \quad \ u_\rom{f}(0)=0. \label{a.2n}
\eequ 
Furthermore, using (\ref{2.5}) and (\ref{2.2}), the time derivative of the error signal is written in the form given by (\ref{a.3}). 
Once again, using (\ref{total:unc:3}) and (\ref{prm:1}) in (\ref{a.3}), we arrive
\bequ 
    \dot{e}(t)&\hspace{-0.25cm}=\hspace{-0.25cm}&A_\rom{n}e(t)+B\Lambda\bigl(u_\rom{f}(t)-\tilde{W}\mT(t)\sigma(x(t),u_\rom{n}(t)) \nonumber\\&& +\epsilon(x(t))  \bigl), \quad e(0)=0. \label{a.4n}
\eequ 
We are now ready to present our second main result, where recall that $\tilde{W}(t)\triangleq \hat{W}(t)-W$ and $\tilde{\Lambda}(t)\triangleq \hat{\Lambda}(t)-\Lambda$. 

\vspace{0.1cm}

\textbf{Theorem 3.} Consider the dynamical system given by (\ref{2.5}) with the nonparametric uncertainty given by (\ref{non:prm:unc}). 
In addition, consider the form of the fixed-gain control signal given by (\ref{ca1.1}) with (\ref{ca1.2}) and (\ref{ca2.3}). 
Consider also the form of the adaptive learning signal given by (\ref{prm:1}), (\ref{ca2.4}), and (\ref{2.3}). 
If $\rho\neq 0$ in (\ref{fc:1}), the trajectories $\bigl(e(t),u_\rom{f}(t),\tilde{W}(t),\tilde{\Lambda}(t)\bigl)$ of the closed-loop system are then bounded according to
\bequ
    \mathcal{V}(\cdot) &\hspace{-0.25cm}=\hspace{-0.25cm}& \beta_1 \kappa \bigl(e\mT(t) P e(t)\bigl)+\beta_2 u_\rom{f}\mT(t)u_\rom{f}(t)\nonumber\\&&
+\rom{tr} \bigl(\tilde{W}(t)\Lambda^{\frac{1}{2}}\bigl)\mT  \bigl(\tilde{W}(t)\Lambda^{\frac{1}{2}}\bigl)\nonumber\\&&
+\beta_1\gamma_1^{-1}\rom{tr}\tilde{\Lambda}\mT (t) \tilde{\Lambda}(t)\ \le \ \mathcal{V}^\star, \label{sp:01} \\
\mathcal{V}^\star &\hspace{-0.25cm}\triangleq\hspace{-0.25cm}& \beta_1 \overline{\lambda}(P)l_1^{-1}l_5+\beta_2l_2^{-1}l_5+\overline{\lambda}(\Lambda)l_3^{-1}l_5\nonumber\\&&+\beta_1\gamma_1^{-1}l_4^{-1}l_5, \label{here:01}
\eequ 
where $l_1\triangleq \beta_1(\underline{\lambda}(R)-d_1)\underline{\kappa}'$ with 
$d_1\in(0,\underline{\lambda}(R))$ and $\underline{\kappa}'\triangleq \rom{min}_{e\mT(t)Pe(t)}\kappa'\bigl(e\mT(t)Pe(t)\bigl)\in\IR_+$, 
$l_2\triangleq (2-d_2) \alpha \beta_2 \underline{\lambda}(\Lambda)$ with $d_2\in(0,2)$, 
$l_3 \triangleq (2-d_3)\beta_3 \underline{\lambda}(\Lambda)$ with $d_3\in(0,2)$, 
$l_4 \triangleq (2-d_4)\beta_1 \gamma_1^{-1} \gamma_2$ with $d_4\in(0,2)$, and 
$l_5 \triangleq \beta_1 d_1^{-1}||P||^2_2 ||B||_2^2 ||\Lambda||_2^2\overline{\epsilon}^2+\alpha\beta_2d_2^{-1}||\Lambda||_2 \overline{\epsilon}^2+\beta_3d_3^{-1}||W||_\rom{F}^2\overline{\lambda}(\Lambda)+\beta_1\gamma_1^{-1}\gamma_2d_4^{-1}\overline{\lambda}^2(\Lambda)$ with $\overline{\kappa}'\triangleq \rom{max}_{e\mT(t)Pe(t)}\kappa'\bigl(e\mT(t)Pe(t)\bigl)\in\IR_+$ and $||\epsilon(x(t))||_2 \le \overline{\epsilon}\in\IR_+$. 

\vspace{0.1cm}

\textit{Proof.} Consider the energy function given by (\ref{sp:01}). 
Note that $\mathcal{V}(0,0,0,0)=0$, $\mathcal{V}(e(t),u_\rom{f}(t),\tilde{W}(t),\tilde{\Lambda}(t))\in\IR_+$, and $\mathcal{V}(e(t),u_\rom{f}(t),\tilde{W}(t),\tilde{\Lambda}(t))$ is radially unbounded. 
The time derivative of (\ref{sp:01}) satisfies\footnote{The same steps taken in the proof of Theorem 2 are not repeated.}
\vspace{0cm}
\bequ 
    \dot{\mathcal{V}}(\cdot) 
    &\hspace{-0.25cm}=\hspace{-0.25cm}& -\beta_1\kappa'(\cdot)e\mT(t)R e(t)-2\alpha \beta_2 u_\rom{f}\mT(t)\Lambda u_\rom{f}(t) \nonumber\\
    &&+2\beta_1 \kappa'(\cdot)e\mT(t) P B \Lambda \epsilon(x(t))\hspace{-0.05cm}-\hspace{-0.05cm}2\alpha \beta_2 u\mT_\rom{f}(t)\Lambda \epsilon(x(t))\nonumber\\
    && -2\beta_3\rom{tr}\tilde{W}\mT(t)\bigl( \tilde{W}(t)+W \bigl)\Lambda -2\beta_1\gamma_1^{-1}\gamma_2\rom{tr}\tilde{\Lambda}\mT(t)\nonumber\\
    &&\times \bigl(\tilde{\Lambda}(t)+\Lambda  \bigl).
    \label{sp:02}
\eequ 
Next, we resort to Young's inequality for sign-indefinite terms in (\ref{sp:02}) as 
\vspace{0cm}
\bequ
2 e\mT(t) PB \Lambda \epsilon(x(t)) &\hspace{-0.25cm}\le\hspace{-0.25cm}& d_1 \beta_1 \kappa'(\cdot)e\mT(t)e(t) \nonumber\\&& +\beta_1 \overline{\kappa}' d_1^{-1}||P||_2^2||B||_2^2||\Lambda||_2^2 \overline{\epsilon}^2, \ \ \ \ \ \label{in:eq:1} \\
2 u_\rom{f}\mT(t)\Lambda \epsilon(x(t)) &\hspace{-0.25cm}\le\hspace{-0.25cm}& d_2 u_\rom{f}\mT(t) \Lambda u_\rom{f}(t) +d_2^{-1}||\Lambda||_2\overline{\epsilon}^2, \label{in:eq:2} \\
2\rom{tr}\tilde{W}\mT(t)W\Lambda &\hspace{-0.25cm}\le\hspace{-0.25cm}& d_3 \rom{tr}\tilde{W}\mT(t)\tilde{W}(t)\Lambda \nonumber\\&& +d_3^{-1}||W||_\rom{F}^2 \overline{\lambda}(\Lambda), \label{in:eq:3}  \\
2\rom{tr}\tilde{\Lambda}\mT(t)\Lambda &\hspace{-0.25cm}\le\hspace{-0.25cm}& d_4 \rom{tr}\tilde{\Lambda}\mT(t)\tilde{\Lambda}(t)+d_4^{-1}\overline{\lambda}^2(\Lambda).  \label{in:eq:4} 
\eequ 
Using (\ref{in:eq:1}), (\ref{in:eq:2}), (\ref{in:eq:3}), and (\ref{in:eq:4}) in (\ref{sp:02}) results in 
\vspace{0cm}
\bequ 
    \dot{\mathcal{V}}(\cdot) 
    &\hspace{-0.25cm}\le \hspace{-0.25cm}& -l_1 ||e(t)||^2_2-l_2 ||u_\rom{f}(t)||_2^2-l_3\rom{tr}\tilde{W}\mT(t)\tilde{W}(t)\nonumber\\&&
    -l_4 \rom{tr}\tilde{\Lambda}\mT(t)\tilde{\Lambda}(t)+l_5. 
    \label{sp:03}
\eequ 
Now, one can immediately observe that (\ref{sp:03}) implies $\dot{\mathcal{V}}(\cdot) \le 0$ outside the compact set {\footnotesize{$\mathcal{S}\triangleq \bigl\{ \xi(t): ||e(t)||_2\le \sqrt{l_1^{-1}l_5} \bigl\} \cap \bigl\{ \xi(t): ||u_\rom{f}(t)||_2\le \sqrt{l_2^{-1}l_5} \bigl\} \cap \bigl\{ \xi(t): ||\tilde{W}(t)||_\rom{F}\le \sqrt{l_3^{-1}l_5} \bigl\}  \cap \bigl\{ \xi(t): ||\tilde{\Lambda}(t)||_\rom{F}\le \sqrt{l_4^{-1}l_5} \bigl\}$}}, where $\xi(t)\triangleq (e(t),u_\rom{f}(t),\tilde{W}(t),\tilde{\Lambda}(t))$. 
Therefore, the evolution of ${\mathcal{V}}(\cdot)$ is upper bounded by {\small{$\mathcal{V}(\cdot) \le  \rom{max}_{\bigl(e(t),u_\rom{f}(t),\tilde{W}(t),\tilde{\Lambda}(t)\bigl)\in \mathcal{S}} \mathcal{V}(\cdot) = \mathcal{V}^\star$}} since ${\mathcal{V}}(\cdot)$ cannot grow outside $\mathcal{S}$. 
This implies that the closed-loop system trajectories $\bigl(e(t),u_\rom{f}(t),\tilde{W}(t),\tilde{\Lambda}(t)\bigl)$ are bounded by (\ref{here:01}). 
\hfill $\blacksquare$

\vspace{0.1cm}

\textbf{Remark 6.} We would like to highlight three points. 
First, as opposed to Theorem 2, $\rho=0$ is not permitted in Theorem 3. 
Second, smaller leakage parameters (i.e., $\gamma_2$ in (\ref{ca2.3}) and $\beta_3$ in (\ref{ca2.4})) and larger number of radial basis functions tend to make the upper bound of $\mathcal{V}(\cdot)$ in  (\ref{here:01}) smaller. 
Third, from Theorem 1, an alternative and effective way to stay close to the nominal (i.e., ideal) closed-loop system behavior given by (\ref{2.3}) against high neural network approximation errors and high leakage parameters is to increase $\alpha$ appearing in (\ref{ca1.1}). 

\vspace{0.1cm}

\section{Illustrative Numerical Examples}\label{numerical_examples}

We now present two examples to show the efficacy of the contributions presented in Sections \ref{parametric_uncertainty} and \ref{nonparametric_uncertainty}, where ${\footnotesize{A=\begin{bmatrix}
0 & 1\\
0 & 0
\end{bmatrix}}}$ and ${\footnotesize{B=\begin{bmatrix}
0\\
1
\end{bmatrix}}}$ are considered for (\ref{2.1}); ${\footnotesize{K_1=\begin{bmatrix}
0.16 & 0.57
\end{bmatrix}}}$, ${\footnotesize{K_2=\begin{bmatrix}
0.16
\end{bmatrix}}}$, and a filtered square-wave reference signal are considered for (\ref{2.2}); and $R=I$ is considered for (\ref{L:E}). 
We also consider zero initial conditions. 


\subsection{Parametric Uncertainty Example} \label{example_1}

\begin{figure}[t!] \vspace{-0.0cm}
\hspace{0.3cm}\includegraphics[width=0.44\textwidth]{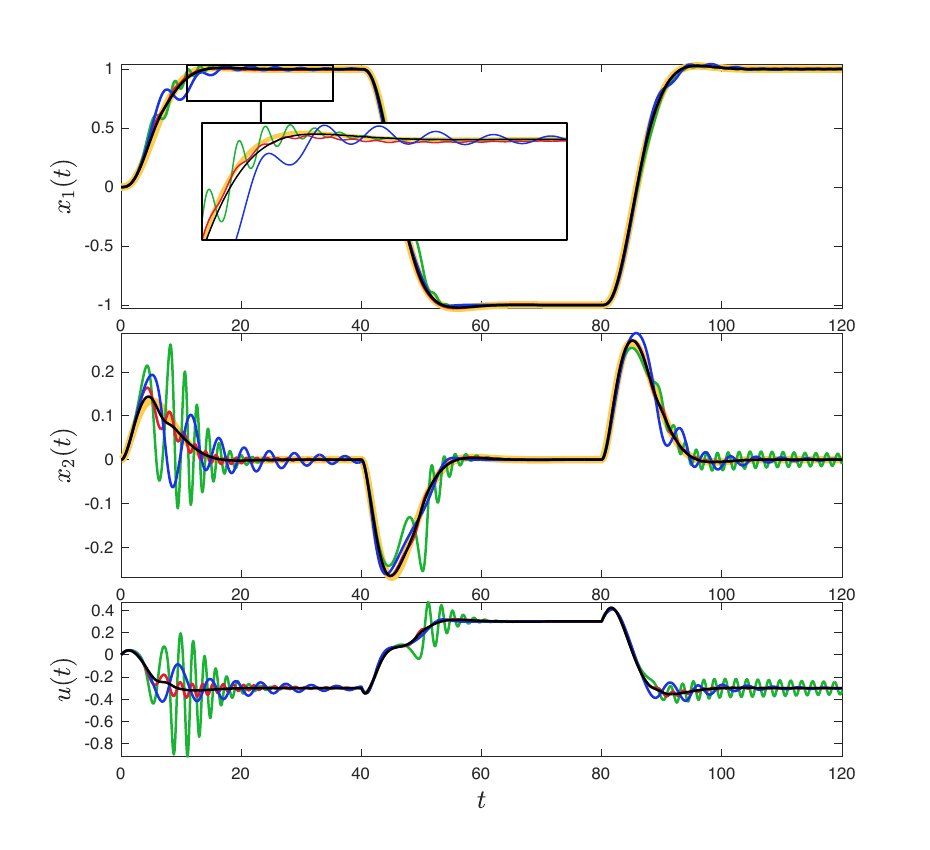} \vspace{-0.7cm}
    \caption{Closed-loop system behavior under parametric uncertainty.} \vspace{-0.6cm}
    \label{results:adaptive} 
\end{figure}

Consider the unknown term $\Lambda=0.9$ representing a $10\%$ degradation in control effectiveness 
and the uncertainty $\delta(x(t))=0.2x_1(t)+0.2x_2(t)+0.8x_1(t)x_2(t)+0.1x_1^3(t)+0.1x_2^2(t)$ for (\ref{2.1}), where this uncertainty is treated as parametric (i.e., $\sigma_\delta(x(t))$ in (\ref{parameterized:unc:here}) is known). 
Figure \ref{results:adaptive} shows the results, where the thick (yellow) line denotes the nominal (i.e., ideal) closed-loop system behavior and the thin (green, red, blue, and black) lines denote the actual closed-loop system behavior for four different cases. 

In particular, the green line denotes the closed-loop system behavior with standard adaptive learning signal and without fixed-gain control signal (i.e., Proposition 2 with $\beta=1$), where state and control responses include oscillations. 
The red line denotes the closed-loop system behavior with symbiotic control signal when $\kappa\bigl(e\mT(t) P e(t)\bigl)=e\mT(t) P e(t)$ and $\kappa'\bigl(e\mT(t) P e(t)\bigl)=1$ (i.e., Theorem 2 with $\alpha=1$, $\beta_1=1$, $\beta_2=1$, and $\gamma=1$), where state and control responses include less oscillations as compared with the former case. 
The blue line denotes the closed-loop system behavior also with symbiotic control signal when $\kappa\bigl(e\mT(t) P e(t)\bigl)$ is chosen as shown in Figure \ref{composite:function} (i.e., Theorem 2 with $\alpha=1$, $\beta_1=1$, $\beta_2=1$, and $\gamma=1$), where state and control responses include oscillations having lower-frequency as compared with the former case. 
The black line denotes the closed-loop system behavior once again with symbiotic control signal when $\kappa\bigl(e\mT(t) P e(t)\bigl)$ is chosen as shown in Figure \ref{composite:function} (i.e., Theorem 2 with $\alpha=3$, $\beta_1=1$, $\beta_2=1$, and $\gamma=1$), where the fixed-gain control parameter $\alpha$ is increased motivated by Theorem 1 to achieve a smooth closed-loop system behavior that remains sufficiently close to its nominal one. 

\subsection{Nonparametric Uncertainty Example} \label{example_2}

\begin{figure}[t!] \vspace{-0.0cm}
\hspace{0.3cm}\includegraphics[width=0.44\textwidth]{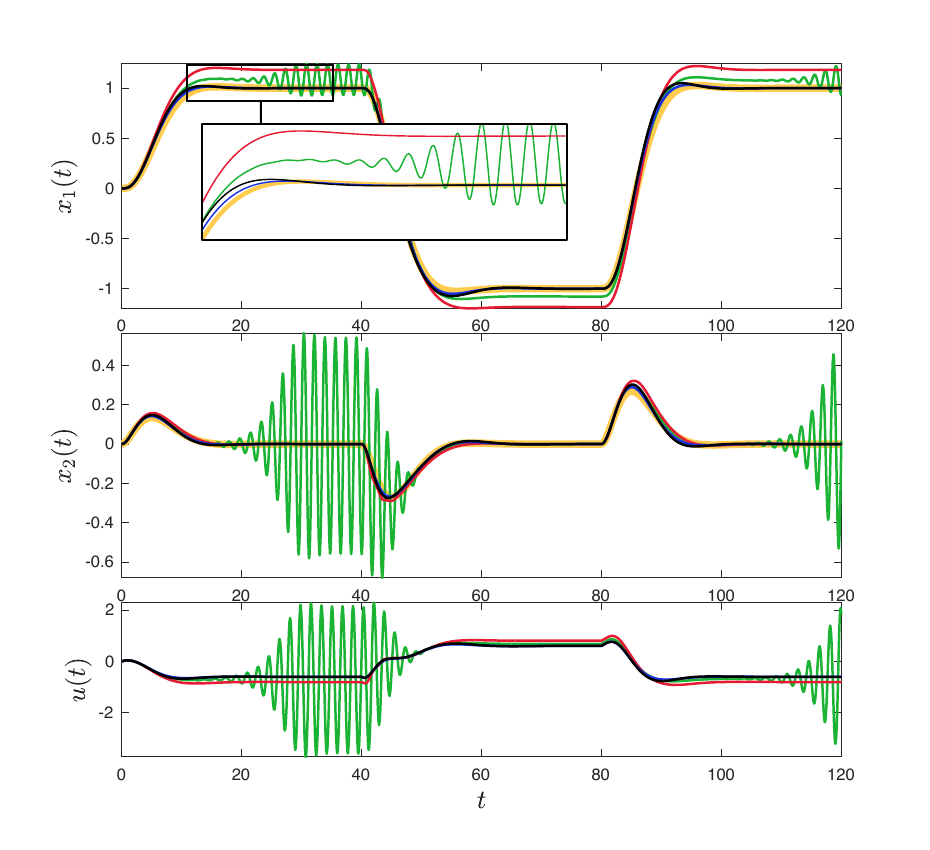} \vspace{-0.7cm}
    \caption{Closed-loop system behavior under nonparametric uncertainty.} \vspace{-0.6cm}
    \label{results:neuroadaptive} 
\end{figure}

Consider the unknown term $\Lambda=0.8$ representing a $20\%$ degradation in control effectiveness 
and the uncertainty $\delta(x(t))=0.4x_1(t)+0.4x_2(t)+1.6x_1(t)x_2(t)+0.2x_1^3(t)+0.2x_2^2(t)$ for (\ref{2.1}), where this uncertainty is treated as nonparametric (i.e., $\sigma_\delta(x(t))$ in (\ref{non:prm:unc}) is constructed with a unity bias and 4 radial basis functions over $\mathcal{D}=(-4,4)\times (-4,4)$ according to $\sigma_\delta(x(t))=$ ${\footnotesize{[1,}}$ {\footnotesize{$\rom{e}^{-0.5(x_1(t)-1)^2},\rom{e}^{-0.5(x_1(t)+1)^2},\rom{e}^{-0.5(x_2(t)-1)^2},\rom{e}^{-0.5(x_2(t)+1)^2}]\mT$}}). 
Figure \ref{results:neuroadaptive} shows the results, where the thick (yellow) line denotes the nominal (i.e., ideal) closed-loop system behavior and the thin (green, red, blue, and black) lines denote the actual closed-loop system behavior for four different cases. 

In particular, the green line denotes the closed-loop system behavior with standard adaptive learning signal and without fixed-gain control signal (i.e., Proposition 3 with $\beta_1=1$ and $\beta_2=1$), where state and control responses include oscillations. 
The red line denotes the closed-loop system behavior also with standard adaptive learning signal and without fixed-gain control signal (i.e., Proposition 3 with $\beta_1=1$ and $\beta_2=2$), where the leakage term parameter $\beta_2$ is increased to achieve a smooth closed-loop system behavior. 
However, the resulting behavior deviates more from the nominal closed-loop system behavior due to the neural network approximation error and high leakage term parameter. 
The blue line denotes the closed-loop system behavior with symbiotic control signal when $\kappa\bigl(e\mT(t) P e(t)\bigl)$ is chosen as shown in Figure \ref{composite:function} (i.e., Theorem 3 with $\alpha=3$, $\beta_1=1$, $\beta_2=1$, $\beta_3=2$, $\gamma_1=1$, and $\gamma_2=2$), where state responses remain sufficiently close to their ideal ones. 
This shows that the proposed symbiotic framework has the ability to achieve a desired level of closed-loop system behavior even in the presence of neural network approximation errors and high leakage term parameters. 
The black line denotes the closed-loop system behavior with fixed-gain control signal and without any adaptive learning signal (i.e., Proposition 1 with $\alpha=9$), where $\alpha$ is increased to maintain a response similar to the former case\footnote{While this closed-loop system behavior is remarkable without an adaptive learning signal, we may not know in practice how large $\alpha$ needs to be as noted in Remark 1, and therefore, we include this result here only to support Proposition 1.}. 

\section{Conclusion}\label{conclusion}

The contribution of this paper was to present the symbiotic control framework that offered the strengths of fixed-gain control and adaptive learning architectures. 
Specifically, a more predictable closed-loop system behavior was shown to be achieved owing to the presence of the fixed-gain control signal (see Theorem 1). 
Furthermore, the presence of the adaptive learning signal avoided a specific uncertainty structure and an upper bound on this uncertainty, where both parametric (see Theorem 2) and nonparametric (see Theorem 3) uncertainty cases were considered. 
In addition to the presented system-theoretical results, illustrative numerical examples were also given to show that the proposed symbiotic control framework had the ability to achieve a desired-level of closed-loop system behavior even with an insufficient number of neurons and without a deep neural network method, or in the face of high leakage term parameters. 





\bibliographystyle{IEEEtran} 
\baselineskip 12pt
\bibliography{refs.bib}
\end{document}